\documentclass[subeqn,12pt,a4paper]{article}

\usepackage{amssymb,amsmath,amsfonts,amsthm, amscd, mathrsfs,helvet, mathtools}

\usepackage{bm, stmaryrd}
\usepackage{graphicx,verbatim}
\usepackage[usenames, dvipsnames]{color}
\usepackage{psfrag}
\usepackage[all]{xy}

\theoremstyle{plain}

\newtheorem*{theorem*}{Theorem}

\newtheorem*{proposition*}{Proposition}

\newcommand{\tensor}[1]{{\bf \underline{#1}}}

\definecolor{brightBlue}{rgb}{0,0,1}

\definecolor{Violet}{rgb}{0.47,0,1}

 \DeclareMathOperator{\str}{Str}

\def\g{\mathfrak{g}}
\def\h{\mathfrak{h}}

\def\ha{\mbox{\small $\frac{1}{2}$}}
\def\qa{\mbox{\small $\frac{1}{4}$}}

\def\M{\mathcal{M}}

\def\1{\tensor{1}}
\def\2{\tensor{2}}
\def\3{\tensor{3}}
\def\4{\tensor{4}}

\numberwithin{equation}{section}

\renewcommand{\L}[0]{\mathcal{L}}
\def\dss{ {\delta_{\sigma\sigma'}} }
\def\pdss{ {\partial_\sigma\delta_{\sigma\sigma'}} }

\def\beq{\begin{equation}}
\def\eeq{\end{equation}}
\def\beqz{\begin{equation*}}
\def\eeqz{\end{equation*}}
\def\bea{\begin{eqnarray}}
\def\eea{\end{eqnarray}}
\def\nn{\nonumber}

\def\f{\mathfrak{f}}

\def\g{\mathfrak{g}}

\def\ads{$AdS_5 \times S^5$ }

\def\jbp{{\mathcal{J}_L}}
\def\jbm{{\mathcal{J}_R}}
\def\jb{{\mathcal{J}}}

\begin{document}

\begin{center}
\vspace*{2em}
{\large\bf
A lattice Poisson algebra for the Pohlmeyer\\
\vspace{1mm}
reduction of the \ads superstring}\\
\vspace{1.5em}
F. Delduc$\,{}^1$,  M. Magro$\,{}^1$, B. Vicedo$\,{}^2$

\vspace{1em}
\begingroup\itshape
{\it 1) Laboratoire de Physique, ENS Lyon
et CNRS UMR 5672, Universit\'e de Lyon,}\\
{\it 46, all\'ee d'Italie, 69364 LYON Cedex 07, France}\\
\vspace{1em}
{\it 2) Department of Mathematics, University of York,}\\
{\it Heslington, York, YO10 5DD, United Kingdom }
\par\endgroup
\vspace{1em}
\begingroup\ttfamily
Francois.Delduc@ens-lyon.fr, Marc.Magro@ens-lyon.fr, Benoit.Vicedo@gmail.com
\par\endgroup
\vspace{1.5em}
\end{center}

\paragraph{Abstract.}

The Poisson algebra of the Lax matrix associated with the 
Pohlmeyer reduction of the \ads superstring is computed from first principles.
The resulting non-ultralocality is mild, which enables to write
down a corresponding lattice Poisson algebra.

\section{Introduction}

We recently showed in \cite{Delduc:2012qb} that the Poisson algebra of the Lax 
matrix associated with symmetric space sine-Gordon models, defined through a
gauged Wess-Zumino-Witten action with an integrable potential
\cite{Bakas:1995bm}, admits an integrable lattice discretization.
In the present letter we compute the $r/s$-matrix structure 
\cite{Maillet:1985ek} associated with the Pohlmeyer reduction of \ads 
superstring theory \cite{Grigoriev:2007bu,Mikhailov:2007xr} directly from its
representation in terms of a    
 fermionic
 extension of a 
gauged WZW action with an integrable potential. 
We similarly find that it is precisely of the type which,  after
regularization 
as in \cite{SemenovTianShansky:1995ha},
 admits an integrable lattice discretization of the general
form identified in  \cite{Freidel:1991jx,Freidel:1991jv}.

\section{Canonical analysis and Hamiltonian}

To begin with we  briefly  recall some usual notations. We refer the reader 
to \cite{Grigoriev:2007bu} for more details concerning this setup. 
The superalgebra $\f = \mathfrak{psu}(2,2|4)$ admits a $\mathbb{Z}_4$-grading,
$\f = \f^{(0)} \oplus \f^{(1)} \oplus \f^{(2)} \oplus \f^{(3)}$
where $\g = \f^{(0)} = \mathfrak{so}(4,1) \oplus
\mathfrak{so}(5)$. Let $G$ denote the corresponding Lie group. 
The supertrace is compatible with the $\mathbb{Z}_4$-grading,
in the sense that $\str(A^{(m)} B^{(n)}) = 0$ 
for $m+n\neq 0$ mod $4$. The reduced theory relies on the element 
$T= \frac{i}{2} \text{diag}(1,1,-1,-1,1,1,-1,-1) \in \mathfrak{f}^{(2)}$.
It defines a $\mathbb{Z}_2$-grading of $\f$ with $\f^{[0]} = \text{Ker} (\text{Ad}_T)$
and $\f^{[1]} = \text{Im} (\text{Ad}_T)$. Elements   
of $\mathfrak{f}^{[0]}$ 
commute with $T$ while those of $\mathfrak{f}^{[1]}$ anti-commute with $T$
and we have $\str(A^{[0]} B^{[1]}) =0$. Finally, 
projectors on $\mathfrak{f}^{[0]}$ and $\mathfrak{f}^{[1]}$ 
are given respectively by  
$P^{[0]}  = - [T,[T,\,\cdot\,]_+]_+$ and $P^{[1]} = - [T,[T,\,\cdot\,]]$.
Let $\h = \g^{[0]}$ be the subalgebra in $\g$ of elements
commuting with $T$. 
The corresponding Lie group $H$ is  $[SU(2)]^4$.  

\medskip

Our starting point is the field theory introduced in \cite{Grigoriev:2007bu}. 
It corresponds to a fermionic extension of a 
$G/H$ gauged WZW with a potential term. The action we start
with is, taking $\epsilon^{\tau \sigma \xi} = 1$,
\begin{align*} 
\mathcal{S} &=  \ha \int d\tau d\sigma \str(g^{-1} \partial_+ g g^{-1} \partial_- g)
+ \mbox{\small $\frac{1}{3}$} \int d\tau d\sigma d\xi \epsilon^{\alpha \beta \gamma} 
\str( g^{-1} \partial_\alpha g g^{-1} \partial_\beta g g^{-1} \partial_\gamma g)
\nn\\ 
&\qquad - \int d\tau d\sigma \str(A_+ \partial_- g g^{-1} - A_- g^{-1}
\partial_+ g+ g^{-1} A_+ g A_- - A_+ A_-)  \nn \\
&\qquad + \ha \int d\tau d\sigma  
\str(\psi_L [T, D_+ \psi_L] + \psi_R [T, D_- \psi_R]) \nn\\
&\qquad  + \int d\tau d\sigma \bigl(
 \mu^2 \str(g^{-1}TgT)    +
\mu \str(g^{-1} \psi_L g \psi_R) \bigr). 
\end{align*}
The fields $g$, $\psi_R$, $\psi_L$ and the gauge fields $A_\pm$ respectively
take values in $G$, $\f^{(1)[1]}$, $\f^{(3)[1]}$ and in $\h$. The covariant
derivatives  are   
$D_\pm = \partial_\pm - [A_\pm, ]$ with $\partial_\pm  = \partial_\tau \pm 
\partial_\sigma$. 

Generalizing the analysis of \cite{Bowcock:1988xr} to the case considered here, 
one finds that the phase space is spanned by the
fields $(g,\jbp,A_\pm,P_\pm,\psi_L,\psi_R)$. The field $\jbp$ corresponds 
to the left-invariant WZW current. Alternatively, one can 
use instead the right-invariant current $\jbm$,  related to $\jbp$
by
\beqz  
\jbm = -2 \partial_\sigma g g^{-1} + g \jbp g^{-1}.
\eeqz
The fields $P_\pm$ are the canonical momenta of $A_\pm$. 
The non-vanishing Poisson 
brackets are
\begin{subequations}
\begin{align*}
\{ \jb_{L\1}(\sigma), \jb_{L\2}(\sigma') \} &= [C_{\1\2}^{(00)}, \jb_{L\2}] \dss  
+ 2 C_{\1\2}^{(00)} 
\pdss,\\
\{ \jb_{R\1}(\sigma), \jb_{R\2}(\sigma') \} &= -[C_{\1\2}^{(00)}, \jb_{R\2}] \dss  
- 2 C_{\1\2}^{(00)} 
\pdss,\\
\{ \jb_{L\1}(\sigma), g_{\2}(\sigma') \} &= - g_{\2} C_{\1\2}^{(00)} \dss \\
\{ \jb_{R\1}(\sigma), g_{\2}(\sigma') \} &= - C^{(00)}_{\1\2} g_{\2} \dss \\
\{ A_{\pm\1}(\sigma), P_{\pm\2}(\sigma') \} &= C_{\1\2}^{(00)[00]} \dss,\\
 \{ \psi_{R\1}(\sigma), \psi_{R\2}(\sigma') \} &= \bigl[ T_{\2}, C_{\1\2}^{(13)}
\bigr] \dss,\\
\{ \psi_{L\1}(\sigma), \psi_{L\2}(\sigma') \} &= \bigl[ T_{\2}, C_{\1\2}^{(31)} 
\bigr] \dss.
\end{align*}
\end{subequations}
In these expressions $C_{\1\2}^{(ij)} \in \f^{(i)} \otimes  \f^{(j)}$ are 
the components of the tensor Casimir (see \cite{Magro:2008dv} for its 
properties) in the decomposition $C_{\1\2} =
C_{\1\2}^{(00)} + C_{\1\2}^{(13)} + C_{\1\2}^{(22)} + C_{\1\2}^{(31)}$ with
respect to the $\mathbb{Z}_4$-grading. The component $C_{\1\2}^{(00)[00]}$ is
defined in a similar way relative  to the $\mathbb{Z}_2$-grading.

 The standard analysis shows that
there is a total of four constraints,
\begin{subequations} \label{fourc}
\begin{alignat}{2}
\chi_1 &= P_+, &\qquad \chi_2 &= P_-, \\
\chi_3 &= \jb_R^{[0]} + A_+ - A_- - \ha [\psi_L, [T,\psi_L]],
&\qquad \chi_4 &=  \jb_L^{[0]} + A_+ - A_- + \ha [\psi_R, [T,\psi_R]].
\end{alignat}
\end{subequations}
The extended Hamiltonian, which has weakly vanishing 
Poisson brackets with the constraints \eqref{fourc}, is
\begin{align} \label{Ham}
H&= \int \!\! d\sigma  \Big(  \qa \str\bigl( \jbp^2 + \jbm^2\bigr) 
+ \str\bigl( \jb_R^{[0]}  A_+ - \jb_L^{[0]}  A_-\bigr) + \ha \str\bigl[(A_+ - A_-)^2\bigr] \nn\\
&\qquad - \ha \str\bigl(\psi_L [T, \partial_\sigma \psi_L - [A_+, \psi_L]]\bigr) 
- \ha \str\bigl(\psi_R [T, - \partial_\sigma \psi_R - [A_-, \psi_R]]\bigr)\\
&\qquad\qquad\qquad - \mu^2 \str(g^{-1}TgT) - \mu \str(g^{-1} \psi_L g \psi_R) 
+ v_+ P_+ + v_- P_- + \lambda (\chi_3 -\chi_4)\Big)  \nn
\end{align}
with $v_+ - v_- = \partial_\sigma (A_+ + A_-) -[A_+, A_-]$. The combination 
 $\chi_3-\chi_4$ of the constraints generates a gauge invariance. 

\section{Continuum and lattice Poisson algebras}

Up to a gauge transformation, the equations of motion for the fields  
$(\jbp,  g, \psi_L, \psi_R)$ under the Hamiltonian \eqref{Ham} are
equivalent to the zero curvature equation $\{ \L, H \} = \partial_\sigma \M 
+ [\M, \L]$
for the following Lax connection \cite{Grigoriev:2007bu}
\begin{subequations}
\begin{align}
\label{laxmatrix}
\L(z) &= - \ha \jbp - \ha z  \sqrt{\mu} \psi_R - \ha z^2  \mu T  
+ \ha z^{-1} \sqrt{\mu} g^{-1} \psi_L g + \ha z^{-2}  \mu g^{-1} T g,\\ 
\M(z) &= - \ha \jbp + A_- - \ha z  \sqrt{\mu} \psi_R - \ha z^2  \mu T  
- \ha z^{-1} \sqrt{\mu} g^{-1} \psi_L g - \ha z^{-2}  \mu g^{-1} T g.
\end{align}
\end{subequations}
The field $A_+$ entering the equations appears as an arbitrary element of
$\h$. We now have all the ingredients needed to compute the Poisson
bracket of the
Lax matrix \eqref{laxmatrix}. The result reads
\begin{multline} \label{laxcont}
4 \{ \L_{\1}(z_1), \L_{\2}(z_2) \} = [r_{\1\2}(z_1, z_2),
\L_{\1}(z_1) + \L_{\2}(z_2)] \delta_{\sigma \sigma'}\\
+ [s_{\1\2}(z_1, z_2), \L_{\1}(z_1) - \L_{\2}(z_2)] \delta_{\sigma \sigma'} + 2
s_{\1\2}(z_1, z_2) \partial_{\sigma} \delta_{\sigma \sigma'},
\end{multline}
where the kernels of the $r/s$-matrices are given by
\begin{subequations} \label{kernelsrs}
\begin{align}
r_{\1\2}(z_1, z_2) &= \frac{z_2^4 + z_1^4}{z_2^4 - z_1^4}
C^{(00)}_{\1\2} + \frac{2 z_1 z_2^3}{z_2^4 - z_1^4} C^{(13)}_{\1\2} +
\frac{2 z_1^2 z_2^2}{z_2^4 - z_1^4} C^{(22)}_{\1\2} + \frac{2 z_1^3
z_2}{z_2^4 - z_1^4} C^{(31)}_{\1\2}, \\
 s_{\1\2}(z_1, z_2) &=C^{(00)}_{\1\2}. 
\end{align}
\end{subequations}
One can check explicitly that the kernels \eqref{kernelsrs} coincide exactly
with the ones that would be obtained from the generalization of the alleviation 
procedure proposed in \cite{Delduc:2012qb} to semi-symmetric space $\sigma$-models. This is
simply a matter of replacing the twisted inner product on the 
twisted loop algebra considered in \cite{Vicedo:2010qd} by the
trigonometric one and to compute the corresponding kernels as 
explained in \cite{Delduc:2012qb}.

An important property of the above $r/s$-matrix structure is that $s$ is
simply the projection onto the subalgebra $\g$. In this case, 
the corresponding
Poisson algebra \eqref{laxcont} can be discretized following
\cite{SemenovTianShansky:1995ha} by introducing a skew-symmetric solution
$\alpha \in \text{End}\, \g$ of the modified classical Yang-Baxter equation
on $\g$. Then the matrices
\begin{equation*}
a_{\1\2} = (r + \alpha)_{\1\2}, \qquad b_{\1\2} = (- s - \alpha)_{\1\2}, 
\qquad c_{\1\2} = (- s + \alpha)_{\1\2}, \qquad d_{\1\2} = (r - \alpha)_{\1\2},
\end{equation*}
satisfy all the requirements of
\cite{Freidel:1991jx,Freidel:1991jv} in order to define the following 
consistent lattice algebra,   
\beqz 
4 \{ \L^n_{\1}, \L^m_{\2} \}  =
a_{\1\2} \L^n_{\1} \L^m_{\2} \delta_{m n}
- \L^n_{\1} \L^m_{\2} d_{\1\2}  \delta_{m n} + \L^n_{\1} b_{\1\2} \L^m_{\2}
\delta_{m+1, n}
- \L^m_{\2} c_{\1\2} \L^n_{\1} \delta_{m, n+1}.
\eeqz
This algebra reduces to \eqref{laxcont} in the continuum limit (see  
\cite{Delduc:2012qb}). The corresponding algebra for the monodromy may be 
found in \cite{Delduc:2012qb}.

\section{Conclusion}

We have constructed a quadratic lattice Poisson algebra associated with
the 
fermionic extension of the 
$(SO(4,1) \times SO(5))/[SU(2)]^4$ gauged WZW model with an integrable
potential.
The fact that one is able to write down such a lattice algebra 
is quite 
appealing and 
in sharp contrast with what happens for the canonical Poisson 
structure of the \ads superstring \cite{Magro:2008dv}. Indeed, it brings
  hope of being able to construct a lattice 
quantum algebra related to the 
Pohlmeyer reduction of the \ads superstring.  
The precise link of this Pohlmeyer reduction with the 
 alleviation procedure presented in 
\cite{Delduc:2012qb} is under study.

\paragraph{Acknowledgements} 

We thank J.M. Maillet for useful discussions. B.V. is supported by UK EPSRC
grant EP/H000054/1.

\providecommand{\href}[2]{#2}\begingroup\raggedright\endgroup


\begin{thebibliography}{10}

\bibitem{Delduc:2012qb}
F.~Delduc, M.~Magro, and B.~Vicedo, {\it {Alleviating the non-ultralocality of
  coset sigma models through a generalized Faddeev-Reshetikhin procedure}},
  \href{http://xxx.lanl.gov/abs/1204.0766}{{\tt arXiv:1204.0766}}.

\bibitem{Bakas:1995bm}
I.~Bakas, Q.-H. Park, and H.-J. Shin, {\it {Lagrangian formulation of symmetric
  space sine-Gordon models}},  {\em Phys. Lett.} {\bf B372} (1996) 45--52,
  [\href{http://xxx.lanl.gov/abs/hep-th/9512030}{{\tt hep-th/9512030}}].

\bibitem{Maillet:1985ek}
J.~M. Maillet, {\it {New integrable canonical structures in two-dimensional
  models}},  {\em Nucl. Phys.} {\bf B269} (1986) 54.

\bibitem{Grigoriev:2007bu}
M.~Grigoriev and A.~A. Tseytlin, {\it {Pohlmeyer reduction of AdS$_5$ $\times$
  S$^5$ superstring sigma model}},  {\em Nucl. Phys.} {\bf B800} (2008)
  450--501, [\href{http://xxx.lanl.gov/abs/0711.0155}{{\tt arXiv:0711.0155}}].

\bibitem{Mikhailov:2007xr}
A.~Mikhailov and S.~Sch{\"a}fer-Nameki, {\it {Sine-Gordon-like action for the
  Superstring in $AdS_5 \times S^5$}},  {\em JHEP} {\bf 05} (2008) 075,
  [\href{http://xxx.lanl.gov/abs/0711.0195}{{\tt arXiv:0711.0195}}].

\bibitem{SemenovTianShansky:1995ha}
M.~Semenov-Tian-Shansky and A.~Sevostyanov, {\it {Classical and quantum
  nonultralocal systems on the lattice}},
  \href{http://xxx.lanl.gov/abs/hep-th/9509029}{{\tt hep-th/9509029}}.

\bibitem{Freidel:1991jx}
L.~Freidel and J.~M. Maillet, {\it {Quadratic algebras and integrable
  systems}},  {\em Phys. Lett.} {\bf B262} (1991) 278--284.

\bibitem{Freidel:1991jv}
L.~Freidel and J.~M. Maillet, {\it {On classical and quantum integrable field
  theories associated to Kac-Moody current algebras}},  {\em Phys. Lett.} {\bf
  B263} (1991) 403--410.

\bibitem{Bowcock:1988xr}
P.~Bowcock, {\it {Canonical quantization of the gauged Wess-Zumino model}},
  {\em Nucl. Phys.} {\bf B316} (1989) 80.

\bibitem{Magro:2008dv}
M.~Magro, {\it {The classical exchange algebra of $Ads_5 \times S^5$ string
  theory}},  {\em JHEP} {\bf 0901} (2009) 021,
  [\href{http://xxx.lanl.gov/abs/0810.4136}{{\tt arXiv:0810.4136}}].

\bibitem{Vicedo:2010qd}
B.~Vicedo, {\it {The classical R-matrix of AdS/CFT and its Lie dialgebra
  structure}},  {\em Lett. Math. Phys.} {\bf 95} (2011) 249--274,
  [\href{http://xxx.lanl.gov/abs/1003.1192}{{\tt arXiv:1003.1192}}].

\end{thebibliography}
\end{document}